\documentclass[journal]{IEEEtran}
\usepackage{graphicx}
\usepackage{url}
\usepackage{amsmath}
\usepackage{tikz}
\usepackage{amsmath}

\newcommand \red[1]{#1}

\begin{document}
\title{Control of dual-wavelength laser emission\\ via optical feedback phase tuning}

\author{Robert~Pawlus,~Stefan~Breuer,~and~Martin~Virte % <-this % stops a space
\thanks{R. Pawlus and M. Virte are with the Brussels Photonics Team, Department of Applied Physics and Photonics (TONA/B-PHOT), Vrije Universiteit Brussel, 1050 Brussels, Belgium (robert.pawlus@vub.be, martin.virte@vub.be)}
\thanks{S. Breuer is with the Institute of Applied Physics, Technische Universit\"{a}t Darmstadt, Schlossgartenstraße 7, 64289 Darmstadt, Germany (stefan.breuer@physik.tu-darmstadt.de)}
\thanks{This research was sponsored by the Research Foundation - Flanders (FWO) (grants 1530318N, G0G0319N), the METHUSALEM program of the Flemish Government, the European Research Council (grant 948129 COLOR-UP), the Adolf Messer Foundation (personal grant), the German Research Foundation (DFG) (389193326) and the Open Access Publishing Fund of Technische Universit\"{a}t Darmstadt.}}

\maketitle

\begin{abstract}
We propose and demonstrate a technique to control \red{the balance between} the two amplitudes of a dual-wavelength laser based on a phase-controlled optical feedback. The feedback cavity length is adjusted to achieve a relative phase shift between the desired emission wavelengths, introducing a boost in gain for one wavelength while the other wavelength experiences additional losses. Tuning the optical feedback phase proves to be an effective way to control the gain \& losses, and, thus, to select one or balance the amplitude of the two emission wavelengths. This concept can be easily adapted to any platform, wavelength range and wavelength separations providing that a sufficient carrier coupling and gain can be obtained for each mode. To demonstrate the feasibility and to evaluate the performance of this approach, we have implemented two dual-wavelength lasers with different spectral separations together with individual optical feedback loops onto a InP generic foundry platform emitting around 1550~nm. An electro-optical-phase-modulator is used to tune the feedback phase. With this single control parameter, we successfully achieved extinction ratios of up to 38.6~dB for a 10~nm wavelength separation and up to 49~dB for a 1~nm wavelength separation. 
\end{abstract}

\begin{IEEEkeywords}
Semiconductor Laser, Dual Wavelength, Dual Wavelength Switching, Dual Wavelength Control, optical feedback, photonic integrated circuit, generic foundry platform. 
\end{IEEEkeywords}

\section{Introduction}
\IEEEPARstart{L}{asers} simultaneously emitting at two distinct wavelengths are promising devices for THz or mm-wave generation \cite{Mittleman2018}, \cite{Huang2009} as well as for applications in telecommunication \cite{Ummethala}, \cite{Nagatsuma2016}. While tuning the wavelength separation and power balance between the different wavelength is essential for the sake of versatility and flexibility, the robustness and compactness of the overall laser system are also essential features. Free space solutions based on superimposed light beams or laser sources with external optical feedback are a simple and versatile approach as their wavelengths and relative optical powers can be controlled separately but they result in bulky setups, pose challenging alignment task as well as mechanical and thermal stabilization issues and are not well suited for mass manufacturing \cite{Hoffmann2005, Huang2004}. Naturally, photonic integrated circuits (PICs) appear to be the ideal option to overcome these issues, especially considering the recent boom of generic foundry platforms \cite{Kish2018, Smit2014, Smit2019}. Integrating distinct lasers on the same chip and then merging their beams is a straightforward solution \cite{Osowski1997, Teng2000, Sun2016, Zhao2018, Guzman2021} but the absence of coupling between the different wavelengths prevent their use in some applications, such as two-color self-mixing velocimetry for which the mode competition represent a key element \cite{Gioannini2014a}. \red{Moreover, the phase noises for each laser are uncorrelated and therefore adds up which can be an important drawback \cite{Qi2011}.} An interesting alternative is to use a laser with broadband emission combined with an external forcing mechanism - such as filtered optical feedback - to precisely select and control a few wavelengths. In \cite{Docter2010}, a discretely tunable Fabry-Perot laser based on a four channel optical feedback section to select specific wavelengths was implemented. A similar external control section was implemented in \cite{Ermakov2012, Khoder2013a} to control a ring-laser also capable of emitting on four wavelengths. All of these integrated laser layouts rely on an arrayed-waveguide-grating (AWG) as a wavelength selective element and Semiconductor Optical Amplifier (SOA) gates to select a wavelength and control their amplitude. The design of the AWG itself already poses a challenge and has a large footprint on the PIC. Furthermore, the SOA gates need to be individually controlled and require a balance of several gain currents when simultaneous dual wavelength emission is desired. In short, at least one control parameter per wavelength is required  and careful tuning of multiple parameters is needed to obtain multi-wavelength emission \cite{Khoder2013b}. A similar problem is faced with multi-cavity lasers such as the one proposed in \cite{Carpintero2012} where 16-different cavities have been coupled using an AWG which also plays the role of wavelength selective element.\\
\red{A drawback of the versatile solutions detailed above is that they typically require an active, precise and simultaneous adjustment of multiple control parameters such as the SOA current in each AWG arm in \cite{Docter2010, Ermakov2012, Khoder2013a, Carpintero2012} }. On the other hand, using a single gain section with an intrinsic wavelength selection mechanism tends to improve the multi-wavelength emission robustness, but at the cost of the tunability and versatility. For instance, Quantum-Dot lasers are capable to emit from two separated energy states i.e. two distinct wavelength regions, the ground-state and the excited-state \cite{Markus2003a}. A certain level in control of their emission states can be obtained by tuning the temperature or asymmetrical biasing \cite{Naderi2010a, Wang2007, Breuer2013}, however, these solutions are either relatively slow or require additional external biasing. Another drawback of Quantum-Dot lasers is their replication, as their properties are highly dependent on the dot size, achieving a reliable manufacturing process is challenging. Embedded gratings were proposed to advance towards reliable sources \cite{Naderi2010a} but this appeared to be insufficient to gain complete control over the dual-wavelength emission of the laser. \red{Different solutions were proposed using quantum-well structures including y-shaped cavities with detuned spectrally selective mirrors \cite{Roh2000, Gwaro2020, Mak2021, Huang2021, Koester2022}, such as Distributed Bragg Refletors (DBRs), or engineered reflectors to accomodate emission at two distinct wavelengths \cite{Osborne2007, Zhang2022}.} However, these devices somehow faced similar challenges and limitations in terms of control mechanism.\\
The use of optical feedback has already been investigated to overcome this issue but this was only partially successful. \red{In \cite{Naderi2010a}, optical feedback has been used to effectively trigger emission from the ground state in a Quantum-Dot laser emitting from the excited state. In \cite{Osborne2012a}, a bistable regime triggered by optical feedback in a dual-wavelength quantum-well laser has been uncovered but suppression ratios are not discussed.} In \cite{Virte2014}, a phase sensitive broadband optical feedback was applied onto such a laser and the length of the external feedback cavity was tuned in a sub $\mu m$ range. This lead to a recurring, but limited exchange in optical power between the two states. Further studies revealed the emergence of multiple longitudinal modes within each state to show energy exchanges between each other, being the limiting factor for a full switch between both states \cite{Virte2016}. \\

Achieving stable and controllable dual-wavelength laser emission is a challenge, and it seems particularly hard to overcome all current limitations with a single device. In this work, we propose and successfully demonstrate a simple and efficient method to control a dual-wavelength laser by using an external phase-controlled optical feedback loop. This approach allows to switch efficiently between multiple wavelengths or to balance their output power. The concept is based on optical feedback with the external cavity length adjusted to achieve a relative phase shift of $\pi$ between the two wavelengths when they are coupled back into the laser cavity. This way, we achieve a boost in gain for one wavelength while the other experiences additional losses. An integrated electro-optical-phase-modulator (EOPM) is then used to vary the feedback phase and to tune the gain and losses induced by the feedback for each wavelength. We show that precise control of the laser emission can therefore be achieved using the voltage of the EOPM as the single control parameter. We experimentally demonstrate the feasibility and efficiency of this approach in two different dual-wavelength laser schemes with wavelength separations of 1~nm and 10~nm. 

\section{Principle of operation}
\begin{figure}
	\centering
	\includegraphics[width=\linewidth]{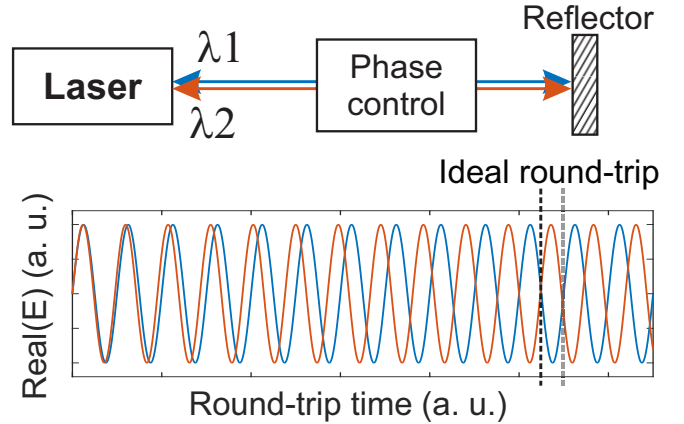}
	\caption{Schematic description of the proposed control approach. A reflector is used to create an external feedback cavity including a component allowing control of the feedback phase. The key parameter is the length of the external cavity as the two wavelengths must ideally be in anti-phase after a round trip in the external cavity. The phase control can then be used to tune the round-trip time at the wavelength scale to select which of the two wavelength is resonating. The dashed black (gray) line shows the ideal mirror position to obtain a resonant blue (red) mode corresponding to $\lambda_1$ ($\lambda_2$).} 
	\label{Fig:Scheme}
\end{figure}

The technique we present in this work relies on a simple system, as shown in Fig. \ref{Fig:Scheme}. Starting with a standalone dual-wavelength laser, it only requires an external cavity, i.e. placing a mirror to partially reflect the light back into the laser cavity. Only part of the emitted light should be fed back to avoid dynamical instabilities \cite{Ohtsubo2013}, but this is not creating any difficulty in practical implementations. The essential point is, however, that the two emitted wavelengths should be in anti-phase, i.e. with a relative phase-shift of $\pi$ between the two fields, when coupled back into the laser cavity. This feature therefore requires the external cavity length to be precisely set with respect to the period of the beating between the two wavelengths. This is crucial as this feature allows to achieve a wavelength selective resonance in the external cavity: the resonating mode will experience a significant gain boost, while the non-resonating mode (in anti-phase) will experience higher losses. This active gain/loss variation is the mechanism that will steer the emission of the dual-wavelength laser towards a balanced or single wavelength output. Finally, a phase controller or modulator placed in the external cavity will provide the necessary tunability of the feedback round-trip time at the wavelength scale. For instance, a phase modulation of $\pi$ would shift the black dashed line, shown in Fig. \ref{Fig:Scheme}, to the position of the gray dashed line, thus moving from a resonant wavelength $\lambda_1$ in blue to a resonant wavelength $\lambda_2$ in red. Naturally, the available range of the phase controller must be sufficient to induce such a shift. A range of $2\pi$ would of course be ideal to compensate any offset that might occur at the manufacturing stage.
\red{It is important to note that, here, we discuss phase control of an \textbf{external} optical feedback, as opposed to phase tuning inside the laser cavity which would impact the longitudinal mode distribution. This is litterally an add-on to a dual-wavelength laser structure and not a modification of the laser itself.}

\section{Theoretical model and simulation results}
To determine the optimal operating condition of the feedback section to achieve the most efficient control of the two wavelengths, we have first investigated our system numerically. We use a multi-mode generalization of the Lang-Kobayashi equations identified as ``Model B'' in \cite{Koryukin2004a} and which reads as follows when considering only two distinct modes: 
\begin{align}
	\frac{dE_1}{dt} = & \frac{1+i\alpha}{2}\left( g_1 N_1 - \frac{1 - g_1}{2} \right)E_1 +\kappa e^{-i \phi_1} E_1\left(t-\tau \right) \label{eq:elec1}\\
	\frac{dE_2}{dt} =  &\frac{1+i\alpha }{2} \left( g_2 N_2 - \frac{1 - g_2}{2} \right)E_2 + \kappa e^{-i \phi_2} E_2 \left(t-\tau \right) \label{eq:elec2}\\
	T \frac{dN_1}{dt} & = P - N_1 - (1 + 2 N_1) \left( g_1 |E_1|^2 + g_2 \beta_{12} |E_2|^2 \right) \label{eq:carrier1}\\
	T \frac{dN_2}{dt} & = P - N_2 - (1 + 2 N_2) \left( g_1 \beta_{21} |E_1|^2 + g_2 | E_2|^2 \right) \label{eq:carrier2}
\end{align}
Equations \ref{eq:elec1} and \ref{eq:elec2} describe the evolution of the two electrical fields $E_1$ and $E_2$ corresponding to the two distinct emission wavelengths $\lambda_1$ and $\lambda_2$, respectively. In both equations, the last term corresponds to the optical feedback $E(t-\tau)$ with the time-delay $\tau$ and feedback rate $\kappa$. Even though the feedback phase $\phi$ is intrinsically linked to the time-delay and the pulsation: $\phi = \omega\tau$, we use two distinct parameters $\phi_1$ and $\phi_2$ that we will set freely to better analyse the impact of the phase difference. Moreover, it has been already seen that a phase variation correspond to a change of the time-delay at such a small scale that the two quantities can, in practice, be considered independently \cite{Ohtsubo2013}. Equations \ref{eq:carrier1} and \ref{eq:carrier2} describe the evolution of the carrier population $N_1$ and $N_2$ with a cross-saturation parameter $\beta_{xy}$, modelling the coupling strength between the two carrier pools. It can be noted that, in these equations, the wavelength of each field only appears in the feedback phase. While the cross-saturation parameters would certainly partially depend on the wavelength, it is interesting to note that this model is virtually independent of the wavelength difference. The model is also normalized in time by the photon lifetime and the time-scale parameters are therefore dimensionless. Unless stated otherwise, we use the following parameters: a symmetrical cross-saturation parameter $\beta_{12} = \beta_{21} = 0.995$, a linewidth enhancement factor of $\alpha = 3$ and a normalized carrier lifetime of $T = 1000$. A low amount of spontaneous emission noise with a noise coefficient of $10^{-20}$ is included to avoid that the model remains on an unstable steady-state. The pump parameter is set at $P=0.5$ which corresponds to an injection current of 2 times the laser threshold. The feedback delay is set at $\tau = 50$ which places us in the short-cavity regime corresponding well to a typical integrated external cavity. The optical feedback phases $\phi_1$ and $\phi_2$ for each wavelength are tuned individually in a range of 0 to 2$\pi$ when the light is fed back into the laser cavity which covers all possible configurations. The gain coefficient will also be slightly varied to evaluate its potential impact on the proposed control technique.\\
Last, to evaluate the switching performance, we define a figure of merit with $FoM = tanh(\frac{P_{1}}{P_{2}}-\frac{P_{2}}{P_{1}})$ with $P_1$ and $P_2$ being the optical power emitted at each wavelength when the laser reaches it's final steady state. This figure gives us a value ranging from $FoM = -1$ - corresponding to $P2 > 0$ and $P1 << P2$ - to $+1$ - corresponding to the inverse situation $P1 > 0$ and $P2 << P1$. A value of $FoM = 0$ correspond to a perfectly balanced output between the two wavelengths, i.e. $P1=P2$. A difference of 3 dB corresponds to $FoM = \pm 0.9$. This figure of merit is only computed when the laser is on a steady-state and not as a time-dependent value. \\

\begin{figure}
	\centering
	\includegraphics[width=\linewidth]{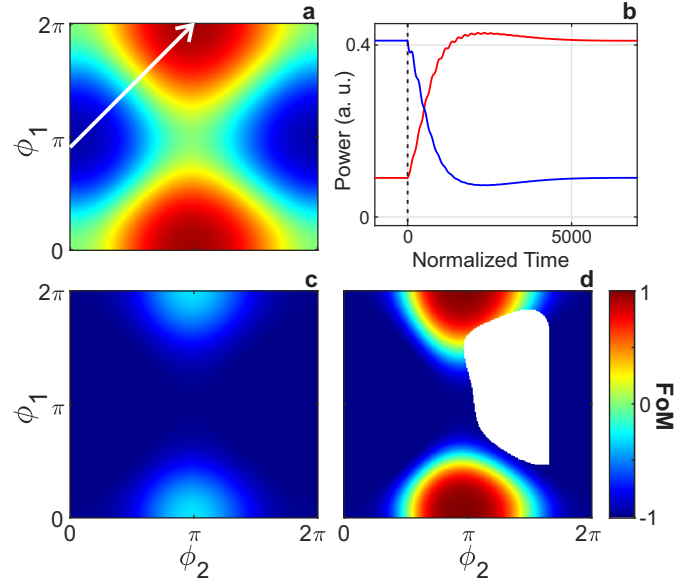}
	\caption{Simulated mappings of the laser behaviour depending on the feedback phase $\phi_{1,2}$. An identical gain $g_1 = g_2 = 1$ for both wavelengths  with a feedback rate of $\kappa = 4 \cdot 10^{-4}$ is considered in (a). The switching behavior induced by the phase change highlighted by the white arrow is given (b), where the vertical dashed line indicate the occurrence of the phase change. The effect of a gain difference $g_1 = 0.998$, $g_2 = 1$ for a feedback rate of $\kappa = 4 \cdot 10^{-4}$ and $10^{-3}$ is shown in (c) and (d) respectively. The color scale indicate the $FoM$ value: a dominant $\lambda_1$ being displayed in red while a dominant $\lambda_2$ appears in blue. An equal output power correspond to $FoM = 0$ and appears in green. Configurations leading to dynamical behaviour are shown in white.}
	\label{FIG: Simulation}
\end{figure}

For a perfectly symmetrical case without feedback, i.e. with $g_1 = g_2 = 1$ and $\kappa=0$, we obtain a balanced simultaneous emission of both modes with $P1=P2$. However, as soon as optical feedback is considered, the equilibrium is perturbed. We observe the emergence of the pattern shown in Fig. \ref{FIG: Simulation}(a) which immediately confirms the significant influence of the feedback phase on the laser emission even at weak feedback levels (here, $\kappa = 4 \cdot 10^{-4}$). While balanced simultaneous emission is maintained when the feedback phase is identical for both modes $\phi_1 = \phi_2$ or inverse $\phi_1 = -\phi_2$ (modulo $2\pi$; this case corresponds to the decreasing diagonal line), one mode becomes dominant when a phase difference is observed. The extreme case appearing when one mode is resonant in the feedback cavity whereas the other is anti-resonant, i.e. $(\phi_1, \phi_2)= (0, \pi)$ or $(\phi_1, \phi_2)= (\pi, 0)$: it is in these configurations that we observe the strongest imbalance between the two modes, with the resonant mode being strongly dominant. A sudden change of the phase configuration - illustrated by the white arrow in Fig. \ref{FIG: Simulation}(a) - leads to a quick switching between the modes as shown in (b). At time $t=0$, the phase configuration is changed leading to the rise of $\lambda_1$ in red accompanied by the suppression of $\lambda_2$ in blue. \\ 
More realistically, a perfect symmetry between the two wavelengths cannot be expected as the two modes will, at the very least, experience different gains. As could be expected, taking a gain asymmetry into account breaks the symmetry of the pattern observed in the feedback phase mapping. Fig. \ref{FIG: Simulation}(c) shows the map evolution for $g_1 = 0.998$ and $g_2 = 1$. Though the gain difference is relatively small, we see that the $FoM$ remains well into the negative range meaning that $\lambda_2$ is the dominant mode for all possible configurations. For $(\phi_1, \phi_2) = (0, \pi)$, we see a slight increase of $FoM$ which confirm that the feedback still has an influence even though it cannot induce a full switch to $\lambda_1$ any longer.\\
Intuitively, the use of a stronger feedback can be expected to amplify the feedback forcing and eventually out balance the gain difference. This assumption is verified in Fig. \ref{FIG: Simulation}(d) which shows the same map but with a higher feedback rate of $\kappa = 0.001$. As expected, with the higher feedback rate, a $FoM$ close to 1 can be obtained around the position $(\phi_1, \phi_2) = (0, \pi)$, which means that a full switch between the two wavelengths can again be achieved. However, this is at the cost of significant dynamical instabilities induced by the feedback and displayed in white. Although optical feedback from short-external cavities are known to induce less instabilities, it is not entirely surprising that, combined with the internal mode competition, it triggers dynamical behaviour. On the bright side, we see that the region where dynamics emerge is rather well circumscribed. On the other hand, we observe that the transition between the two regions showing dominant $\lambda_1$ and $\lambda_2$ emission respectively is narrower. Though it might be an advantage for switching between the two wavelengths, this feature also means that achieving a balanced emission will be more difficult as the feedback phase will need to be adjusted more precisely. \\
Though a detailed investigation of the influence of the different parameters such as the feedback rate, the time-delay or the cross-saturation coefficient on the switching performances is out of the scope of this paper and is left for future work, these simulation results provide key information in the context of controlling dual-wavelength lasers. First, it appears that tuning the feedback phase could indeed be a suitable mean to control the emission of a dual-wavelength laser. Second, we can see that $(\phi_1, \phi_2) = (0, \pi)$ and $(\pi, 0)$ are the two configurations for which the strongest forcing can be achieved for $\lambda_1$ and $\lambda_2$, respectively. But, the maps shown in Fig. \ref{FIG: Simulation} also suggest that this setting does not need to be perfectly accurate to achieve good forcing. Third, it is clear that the emergence of undesired dynamical behaviour due to feedback will be a main concern and will have to be carefully monitored.\\ 
In practice, adding a phase control in the external feedback cavity will change both $\phi_1$ and $\phi_2$ simultaneously. We can reasonably assume that both wavelength will experience a similar phase shift. Thus, experimentally, we will only be able to travel in the phase maps along one diagonal line parallel to the white arrow in Fig. \ref{FIG: Simulation}(a). The position of this line being fixed by the phase difference between the two wavelength emission after a round trip in the external cavity. To achieve both optimal phase configurations highlighted above and move along the white arrow shown in Fig. \ref{FIG: Simulation}, we should therefore design the external cavity such that $\lambda_1$ and $\lambda_2$ are in an anti-phase configuration after a round-trip in the external cavity. This is exactly the reasoning we will implement experimentally on PICs in the next section. 

\section{Implementation on Photonic ICs}
As discussed in the previous section, to implement a demonstrator onto a PIC, the optical feedback cavity needs to be tailored to ideally induce a $\pi$ phase shift between the two wavelengths of the dual-wavelength laser, i.e. the length of the feedback cavity must be set, so that a relative phase shift of $\pi$ between the two wavelengths is achieved after one round trip. To determine this suitable cavity length, we need to know the two emission wavelengths $\lambda_1$ and $\lambda_2$ at the design stage with a good precision. In an external cavity of length $L$, the total round trip length being $L_{FB}=2L$. With effective index $n_{eff}$, the amplitude envelope of the beat-note of the two different wavelengths after a round-trip is given by $2cos(\pi L_{FB} n_{eff} (1/\lambda_1 - 1/\lambda_2))$. $\lambda_1$ and $\lambda_2$ will be in anti-phase when this envelope is reaching zero. As a result, we obtain that the feedback length $L$ should ideally be equal to: 
\begin{equation}
	L_{FB} = (m + 1/2) \frac{\lambda_1 \lambda_2}{n_{eff} (\lambda_1 - \lambda_2)}
\end{equation}
with $m$ a positive integer. Besides the need for good estimates of the effective index and the wavelengths, we can see from this equation that shorter cavities will have an improved robustness against wavelength and index estimation errors. Combined with the improved robustness of semiconductor laser against short-external cavity feedback, one should try to use the shortest external cavity possible. \\ 

\begin{figure}
	\centering
	\includegraphics[width=\linewidth]{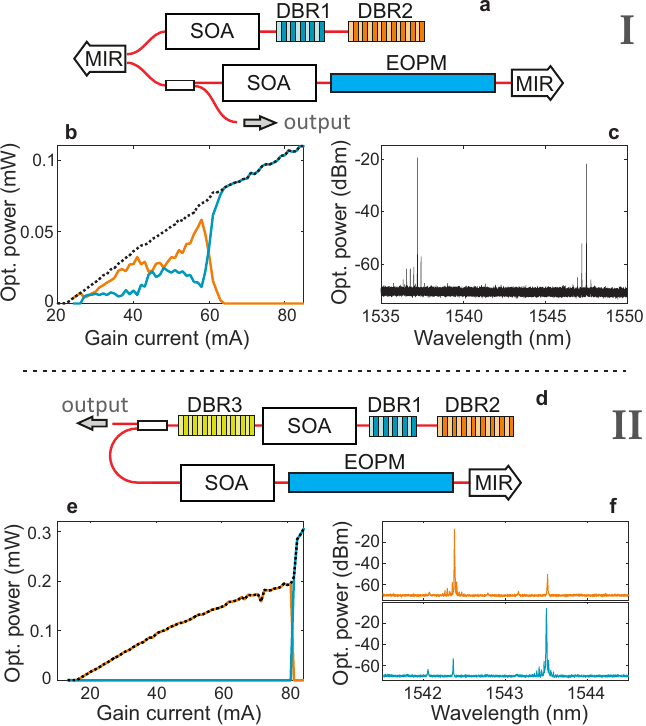}
	\caption{\red{Schematic of the two dual-wavelength lasers. \textbf{Layout I}: (a) schematic of the laser design composed of two sequentially placed DBRs combined with a broadband MIR and coupled to an external cavity via a 1x2 MMI splitter. (b) wavelength resolved LI curve of the laser with $\lambda_1$ ($\lambda_2$) shown in blue (orange) and total power showed as a black dotted line. (c) Optical spectrum of the laser for an injection current of 30 mA.\linebreak \textbf{Layout II}: (d) schematic of the laser design composed of a combination of three distinct DBRs as described in the text, again coupled to an external cavity via a 1x2 MMI splitter. (e) wavelength resolved LI curve with the same color code than for layout I. (f) Optical spectra at 81 (top) and 82 mA (bottom) showing emission at $\lambda_1$ and $\lambda_2$ respectively.}}
	\label{FIG: Laser layouts}
\end{figure}

For the experimental demonstration of our approach, we have designed two dual-wavelength lasers, shown schematically in Fig.~\ref{FIG: Laser layouts} (I) \& (II), with two sequentially arranged Distributed Bragg Reflectors (DBRs) to select the emission wavelengths. \red{These two designs are build using only the building blocks provided in the SMART Photonics library: each element is therefore almost entirely predefined with only a few parameters which can be adjusted. All accessible parameters are provided in the text.} Since the grating coefficients were fixed by the foundry process, only the length of the DBRs were adjusted. We therefore had to make a trade-off between DBR bandwidth and reflectivity, for which we used the INTERCONNECT software from LUMERICAL to fine tune our design. In practice, we made $DBR_1$ shorter - as detailed below - to ensure that the laser cavity formed by $DBR_2$ would also lead to single mode emission. The Semiconductor Optical Amplifier (SOA) used as gain medium is 500 $\mu m$ long in both cases and is followed by a mode filter.\\
To close the laser cavity, we either used a Multimode Interference Reflector (MIR) or a third DBR; both options exhibiti different constraints and advantages. Closing the laser cavity with a MIR, as shown in Fig. \ref{FIG: Laser layouts} (I), does not constrain the wavelength separations. Although it is ultimately limited by the width of the gain spectrum provided by the SOA used as gain medium, the MIR has a relatively flat reflectivity over a large bandwidth which is not a limiting factor in this case. Our design considered a 10~nm splitting between the two wavelengths of our dual-wavelength laser defined by the Bragg Wavelength of each DBR with $\lambda_1 = 1545 \, nm$ and $\lambda_2 = 1535 \, nm$. Their lengths have been set at $250 \, \mu m$ and $500 \, \mu m$ for DBR 1 and 2, respectively. Taking the effective DBR lengths into account, the laser cavity formed by the MIR and $DBR_1$ is $1080 \, \mu m$ while the cavity formed between the MIR and $DBR_2$ is $1400 \, \mu m$.\\
Alternatively, closing the laser cavity with a third DBR, as shown in Fig. \ref{FIG: Laser layouts} (II), limits the wavelength separation to the width of its reflection spectrum, typically of a few nanometers. Indeed, both wavelengths needs to be reflected enough by $DBR_3$, but with the grating coefficient being fixed, we only have a marginal tunability of the DBR bandwidth. One advantage, however, is that the reflectivity can be adjusted to a lower reflectivity than the MIR, leading to an increase in the output power of the laser. In addition, the DBRs can be slightly tuned via current injection towards shorter wavelengths, and thus it gives some flexibility to adjust the relative reflectivity between the two wavelengths. Although limited, this degree of freedom can be used to improve dual-wavelength emission properties and compensate, to some extent, design errors or manufacturing defects. It should be noted that the tunability of the DBRs is not sufficient to gain full control over the emission of the dual-wavelength laser unlike the phase-controlled feedback technique we propose. In this work, we fixed the wavelengths of $DBR_1$ and $DBR_2$ to $1539.3 \, nm$ and $1541.3 \,nm$ with lengths of $200 \, \mu m$ and $350 \, \mu m$ respectively. $DBR_3$ has a length of $250 \, \mu m$ and its wavelength is set at $1540 \, nm$, 0.3 nm shorter then the average between $\lambda_1$ and $\lambda_2$ to fully benefit from the tuning flexibility mentioned earlier. The laser cavity lengths are $880 \, \mu m$ and $1150 \, \mu m$ respectively when taking the effective DBR lengths into account. With these settings, we determined, based on Lumerical INTERCONNECT simulations, that a 1 nm separation between the two emitted wavelength could be expected with $\lambda_1 = 1539.8 \, nm$ and $\lambda_2 = 1540.8 \, nm$ .\\
In both cases, the light emitted by the lasers is split into two parts via a 2x2 85:15 multimode interference (MMI) splitter, $85 \%$ are guided to the edge of the chip to couple the light out for measurements while $15 \%$ are guided to the feedback section. To avoid reflections from the edge of the PIC, an angled facet of 7$^{\circ}$ is implemented together with an anti-reflection coating to reduce the reflections below -40 dB. \\

% FB control
The feedback sections are similar in both designs and only differ in the length of the external cavity to match the phase condition discussed earlier. 
Each feedback section consists of a SOA with a length of 300 $\mu m$ and an EOPM with a length of 1200 $\mu m$ to control the feedback strength and the optical feedback phase respectively. The EOPM is long enough to provide a phase shift of up to $2\pi$ one way, thus corresponding to a feedback phase shift of up to $4\pi$. The light is reflected back towards the laser cavity via a single port MIR. At this stage, we assume that the MIR will reflect both wavelengths in the same manner. The MIR has a reflectivity of approximately 40 \%. Thus, when the SOA amplification compensates exactly the waveguide and EOPM losses and including the 15 \% transmission of the coupler, about 1 \% of the light is reflected back towards the laser cavity.\\
With all these elements combined and the necessary curved waveguides to make the design fit on a single chip, the smallest implemented external cavity has a length of \mbox{2.4 mm}. Although critical, the ideal cavity length is difficult to estimate: for a wavelength separation of 10 nm (resp. \mbox{1 nm}) the beating period is about 70 $\mu m$ (resp. 700 $\mu m$) for the feedback length, i.e. the distance between two suitable feedback lengths. The manufacturing precision is of the order of nanometers and should therefore be largely sufficient in practice, but the effective refractive index for the external cavity can hardly be estimated. The refractive index in the EOPM and the SOA is not directly accessible, will vary when the EOPM and SOA are operated, and are highly subject to manufacturing variabilities. With a cavity of 2.4 mm, for a 10 nm separation (resp. 1 nm separation), an error of 1.5 \% (resp. 7 \%) on the index covers a full beating period. With this uncertainty, we had to take an arbitrary decision at the design stage: we took a refractive index of 3.36 - corresponding to a typical value of the index of the waveguide material - and used it to estimate suitable feedback lengths. Based on previous insights on multimode dual-state emitting QD lasers, we know that the impact of longitudinal modes can be particularly detrimental \cite{Virte2014, Virte2016}, we therefore also include longitudinal modes in our analysis to ensure that they remain approximately all in phase with the main mode. Since the wavelength separation between longitudinal modes is small, this part is expected to be quite robust against any estimation error. In the end, the total cavity lengths have then been set at 2500~$\mu m$ and 2400~$\mu m$ for laser layout I and II respectively and including the effective propagation length inside $DBR_3$ for the latter. \\

% Setup
After manufacturing, the PIC has been glued on a peltier element including a thermistor and all metal pads on chip have been wafer bonded to a PCB board. A lensed fiber is used to couple the light out of the chip towards instruments. A Thorlabs Pro8 system is used to control the laser SOAs, the DBRs and the feedback SOA as well as to set the temperature of the PIC to 20~$^\circ$C. The EOPM voltage was set by an independent DC source (Agilent E3648A). The optical spectra and optical power are measured by a high-resolution optical spectrum analyzer (Apex AP2083A, resolution down to 5 MHz / 40 fm).

\section{Proof-of-concept demonstration}
The PIC manufacturing is not (yet) a perfect process and some unavoidable manufacturing variations or defects can lead to significant changes of the laser behaviour. Although we successfully obtained dual-wavelength emission with both designs, we slightly tuned some DBR wavelengths to optimize the operating point to achieve best performances. However, as discussed in the following section, the proposed scheme is overall quite robust against parameter variations.  \\

\begin{figure}[tb]
	\centering
	\includegraphics[width = \linewidth]{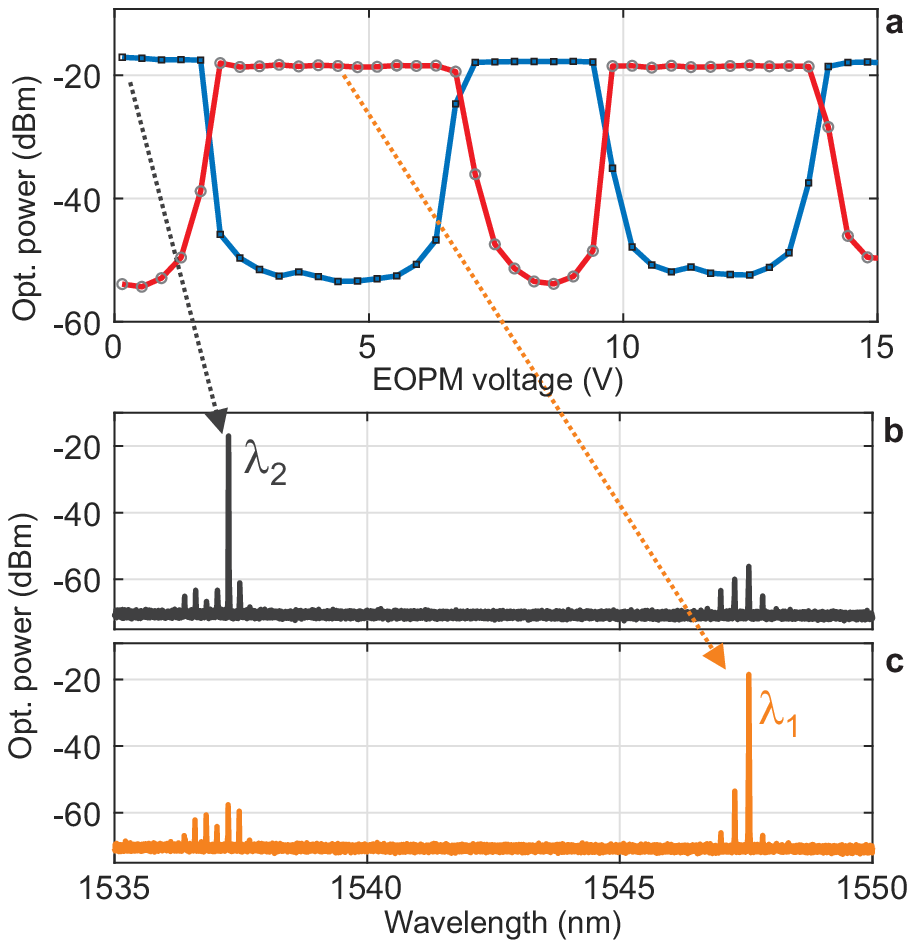}
	\caption{Effect of variations of the EOPM voltage on the laser emission for layout I. (a) Output power evolution for $\lambda_1$ in red line and gray circles and $\lambda_2$ in blue line and black squares for increasing EOPM voltage. (b-c) Optical Spectra for an EOPM voltage of 0V (b) and 4.5V (c) obtained in the same measurement than the data shown in panel (a). Different colors are used only to highlight that the two spectra correspond to different voltages. Experimental conditions are detailed in the text. \label{FIG:XP_layout1}}
\end{figure}

\begin{figure}[tb]
	\centering
	\includegraphics[width = \linewidth]{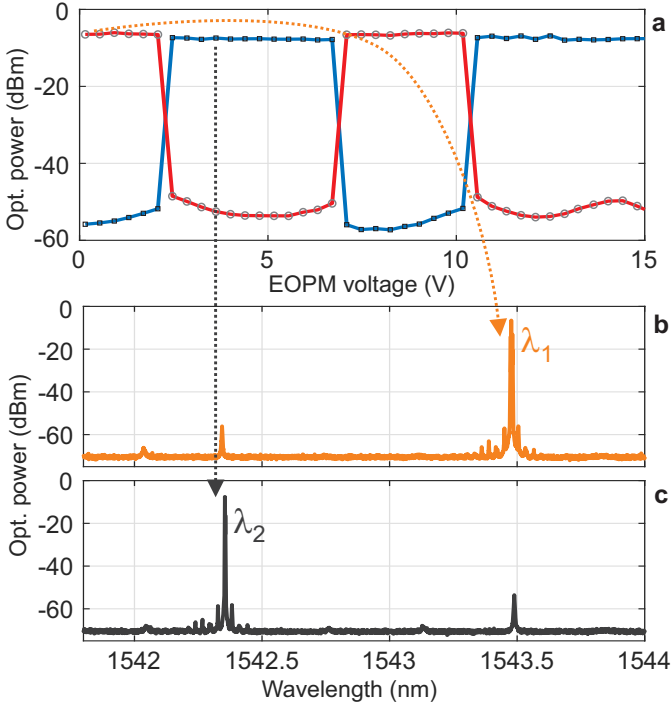}
	\caption{Effect of variations of the EOPM voltage on the laser emission for layout II. (a) Output power evolution for $\lambda_1$ in red line and gray circles and $\lambda_2$ in blue line and black squares for increasing EOPM voltage. (b-c) Optical Spectra for an EOPM voltage of 0 V (b) and 4.5 V (c) obtained in the same measurement than the data shown in panel (a). Different colors are used only to highlight that the two spectra correspond to different voltages. Experimental conditions are detailed in the text. \label{FIG:XP_layout2}}
\end{figure}

\textbf{Layout I}: The laser has a threshold of 22~mA and starts lasing on the shorter wavelength $\lambda_2 = 1537.3 \, ~nm$. Higher currents - starting from 24 mA and above - quickly lead to dual-wavelength emission with multiple switches between the two emission wavelengths as shown in Fig. \ref{FIG: Laser layouts} (b-c). A current of 11~mA is injected into $DBR_1$ to shift the reflection spectrum to shorter wavelengths and to achieve a wavelength separation of $ 10.3 \, nm$ with the longer wavelengths $\lambda_1 = 1547.6 \, nm$. Tuning $DBR_1$ does not lead to other visible changes in the solitary laser behavior. The longitudinal mode separation for $\lambda_1$ is $0.27 \, nm$ and $0.22 \, nm$ for $\lambda_2$. Only a single mode contributes to the emission for each wavelength region while the side modes are well suppressed, typically by at least 30 dB when the corresponding wavelength is dominant. In the following, we use a current of 30 mA for the SOA in the laser cavity as our operating point. The laser spectra without feedback is given in Fig. \ref{FIG: Laser layouts} (c) and a well-balanced simultaneous emission can be observed.\\
When injecting a current of up to 5~mA to the feedback SOA, the laser emission remains unaltered. Above this level, one wavelength starts being favoured with its output power increasing while the other is being suppressed. Although we did not observe dynamics at large currents, stronger feedback did not appear to correlate with a better wavelength selectivity. We actually achieved the best extinction ratio for a rather weak feedback SOA current of 10 mA as shown in Fig.\ref{FIG:XP_layout1}: $\lambda_1$ is suppressed by 37.3 dB compared to $\lambda_2$ at 0 V, while $\lambda_2$ is suppressed by 34.9 dB compared to $\lambda_1$ at 4.8 V, see panels (b) and (c) respectively. The recorded switching behaviour shows a periodicity consistent with the expected EOPM tuning range of $4\pi$ as two full switching cycles appear. We can however note that at 15 V, which is close to the breakdown voltage and well above the recommended operating condition of 8 to 10 V, we can observe a mild degradation of the $\lambda_1$ suppression. \red{Finally, it is interesting to remark that a variation of the EOPM voltage of only a few volts is sufficient to trigger a full switching. Around 7V, the transition is rather smooth, thus suggesting that the same feedback configuration might potentially be used both for switching or adjusting the balance between the two wavelength emissions. Alternatively, using a lower feedback rate by reducing the feedback SOA current reduces the switching amplitude and allows to finely balance the power ratio between the wavelengths.}\\

\textbf{Layout II:} The laser has a lower threshold at 16 mA and starts to emit on the shorter wavelength $\lambda_2 = 1542.4 \, nm$ which is $1.6 \, nm$ longer than the wavelength expected from our simulations. As discussed in the design section, $DBR_3$ has been detuned by 0.3 nm towards shorter wavelength which is consistent with this mode being the first to reach threshold. Tuning $DBR_3$ via injection current appears to be required to achieve emission from the longer wavelength. With a current of $1.7 \, mA$ applied to $DBR_3$, a sudden switching to the longer wavelength with $\lambda_1 = 1543.5 \, nm$ occurs for a laser current around 85.5 mA as shown in Fig. \ref{FIG: Laser layouts} (e-f). This gives a wavelength separation of $1.1 \, nm$ which is well in line with the targeted $1 \, nm$ separation. Increasing current level in $DBR_3$ does not allow to push this switching point to lower laser currents or $\lambda_1$ to higher wavelengths. With this laser, no simultaneous emission is achieved but only sequential emission at the two different wavelengths. \\
Layout II is more responsive to optical feedback due to the lower reflectivity of $DBR_3$ and a feedback current of $6.9 \, mA$ serves as operating point. \red{Turning the optical feedback also seems to induce a shift of the switching point between the two wavelength with respect to the laser current. Thus, we use a laser current of 76.8 mA as operating point for testing the feedback control.} The evolution of the output power of $\lambda_1$ and $\lambda_2$ when varying the EOPM voltage is shown in Fig. \ref{FIG:XP_layout2}. Here, as for layout I, we observe that varying the feedback phase - through the EOPM voltage - allows to effectively select the emitted wavelength. With an EOPM voltage of $0 \, V$, $\lambda_1$ is emitting while $\lambda_2$ is suppressed by 49.7 dB, whereas a voltage of $4.8 \, V$ gives $\lambda_2$ emitting with $\lambda_2$ suppressed by 46.0 dB. Similar to layout I, we observe a degradation of the switching performances for high voltages close to the breakdown voltage: a last switch would have been expected to be consistent with the $4\pi$ phase change, but only a small bump of $\lambda_1$ around 14 V is visible. In addition, we only obtain sharp jumps between the two wavelengths when the EOPM voltage is varied as opposed to the rather smooth transitions obtained for layout I. \red{Surprisingly, we made the same observation even when reducing the feedback strength. Comparison with the results of layout I strongly suggests that this is a device-related issue rather than an issue linked with the optical feedback itself, especially as the same sudden switching is also observed when varying the laser current with feedback.}

\section{Robustness of the proposed technique}
In this paper, we proposed and made a proof-of-concept demonstration of a novel technique to control dual-wavelength lasers via optical feedback. We showed that this approach could be efficiently used to select one emission wavelength with an excellent suppression of the other one or, in some devices, to precisely balance the dual-wavelength emission of the laser. However, the question of the robustness remains to be addressed. \\
\textbf{Dependence on the laser design / type of laser}: In the experimental results shown in the previous section, we observed that the effect of phase-controlled feedback was significantly different between the two type of lasers we considered. The most striking evidence being that with layout II only sudden switching could be achieved while with layout I it was clearly possible to fine tune the power balance between the two wavelengths from equal amplitudes to maximum suppression. Such difference is consistent with the behavior of the two lasers without feedback: layout I showing a smooth transition while only sudden jumps could be recorded with layout II. Yet, it also confirms that the performance of the proposed approach cannot fully overcome all intrinsic limitations of the laser itself. \\
\textbf{Robustness against other control parameters}: For the demonstration, we carefully chose the operating current to be as close as possible to a tipping point between the two wavelengths of the laser. We can naturally assume that this configuration would be the easiest to control. Though a detailed investigation of the robustness of the proposed approach is out of the scope of this paper, our preliminary investigations on this topic tend to confirm that the maximum suppression ratio is reduced when going away from this ideal operating point. Good control - i.e. 10 dB suppression both ways - could still be achieve when the laser injection current was varied by {\protect $\pm$ 10 \%} of the selected operating point. In theory, a stronger feedback might increase the operation range but we did not check this assumption experimentally yet. \red{Alternatively, with the objective of balancing the dual-wavelength emission the operating range will be significantly larger as the required external forcing is much weaker. In this case, however, a weaker feedback might be preferred as it would reduce the switching amplitude and thus improve the control precision though at the cost of reducing the operating range.} \\
\textbf{Influence of the external cavity length}: Although, as mentioned several times in this manuscript, the feedback phase difference between the two wavelengths - and therefore the external cavity length - is the cornerstone of the proposed technique, it is particularly hard to evaluate experimentally the impact of a possible mismatch. The cavity length simply cannot be adjusted on chip and further investigations will require new PICs. However, despite some uncertainty on the selected parameters, convincing experimental results have been achieved. This positive outcome suggests that the external cavity length requirement might not be as stringent as expected and that control of multi-wavelength laser - with more than two distinct wavelengths - could be envisaged, even though further work will still be required to fully confirm this point. 

\section{Conclusion}
We have proposed and experimentally demonstrated a technique to control the amplitude of dual-wavelength lasers using phase-controlled optical feedback. We have shown that this approach offers a simple and effective control mechanism leading to complete switching from one wavelength to the other with a suppression ratio above 30 dB and up to 49 dB. The concept is not limited to a specific wavelength region or platform and can be, in theory, adapted for any wavelengths and wavelength separations meeting the requirements of a similar gain and a sufficient carrier coupling.\\ 
We have implemented two different demonstrators on a generic foundry platform including two distinct dual-wavelength laser designs with 1 and 10 nm wavelength separation. Although emission from either wavelength could be achieved without external forcing, we were able \red{to gain control} of both devices with the proposed phase-controlled optical feedback technique. We showed that varying the EOPM voltage led to repetable switching. \red{The proposed technique is also useful to balance the power ratio between the two wavelengths though it cannot overcome device limitations as could be seen from the impossibility to reach simultaneous emission in devices with layout II.}\\
Last, from a theoretical point of view, we obtain a good quantitative agreement with an extended version of the Lang-Kobayashi model with two coupled emission processes. This suggests that further numerical investigations could provide an important insight to further optimize the proposed approach.\\

\section*{Acknowledgements}
The authors acknowledge the COBRA team from TU Eindhoven for providing us with the COBRA (TUESMARTX) PDK for the SMART Photonics platform. 

\ifCLASSOPTIONcaptionsoff
  \newpage
\fi



%\bibliographystyle{IEEEtran}
\bibliographystyle{IEEEtran}
\bibliography{IEEEabrv,library2, Khoder, Markus, ExtraLib}

\begin{thebibliography}{10}
\providecommand{\url}[1]{#1}
\csname url@samestyle\endcsname
\providecommand{\newblock}{\relax}
\providecommand{\bibinfo}[2]{#2}
\providecommand{\BIBentrySTDinterwordspacing}{\spaceskip=0pt\relax}
\providecommand{\BIBentryALTinterwordstretchfactor}{4}
\providecommand{\BIBentryALTinterwordspacing}{\spaceskip=\fontdimen2\font plus
\BIBentryALTinterwordstretchfactor\fontdimen3\font minus
  \fontdimen4\font\relax}
\providecommand{\BIBforeignlanguage}[2]{{%
\expandafter\ifx\csname l@#1\endcsname\relax
\typeout{** WARNING: IEEEtran.bst: No hyphenation pattern has been}%
\typeout{** loaded for the language `#1'. Using the pattern for}%
\typeout{** the default language instead.}%
\else
\language=\csname l@#1\endcsname
\fi
#2}}
\providecommand{\BIBdecl}{\relax}
\BIBdecl

\bibitem{Mittleman2018}
D.~M. Mittleman, ``{Twenty years of terahertz imaging [Invited]},''
  \emph{Optics Express}, vol.~26, no.~8, p. 9417, 2018.

\bibitem{Huang2009}
J.~Huang, C.~Sun, B.~Xiong, and Y.~Luo, ``{Y-branch integrated dual wavelength
  laser diode for microwave generation by sideband injection locking.}''
  \emph{Optics express}, vol.~17, no.~23, pp. 20\,727--20\,734, 2009.

\bibitem{Ummethala}
S.~Ummethala, T.~Harter, K.~Koehnle, Z.~Li, S.~Muehlbrandt, Y.~Kutuvantavida,
  J.~Kemal, J.~Schaefer, A.~Tessmann, S.~K. Garlapati, A.~Bacher, L.~Hahn,
  M.~Walther, T.~Zwick, S.~Randel, W.~Freude, and C.~Koos, ``{THz-to-optical
  conversion in wireless communications using an ultra-broadband plasmonic
  modulator},'' \emph{Nature Photonics}, 2019.

\bibitem{Nagatsuma2016}
\BIBentryALTinterwordspacing
T.~Nagatsuma, G.~Ducournau, and C.~C. Renaud, ``accelerated by photonics,''
  \emph{Nature Publishing Group}, vol.~10, no.~6, pp. 371--379, 2016. [Online].
  Available: \url{http://dx.doi.org/10.1038/nphoton.2016.65}
\BIBentrySTDinterwordspacing

\bibitem{Hoffmann2005}
S.~Hoffmann, M.~Hofmann, M.~Kira, and S.~W. Koch, ``{Two-colour diode lasers
  for generation of THz radiation},'' \emph{Semiconductor Science and
  Technology}, vol.~20, no.~7, 2005.

\bibitem{Huang2004}
C.~C. Huang, C.~H. Cheng, Y.~S. Su, and C.~F. Lin, ``{174-nm Mode Spacing in
  Dual-Wavelength Semiconductor Laser Using Nonidentical InGaAsP Quantum
  Wells},'' \emph{IEEE Photonics Technology Letters}, vol.~16, no.~2, pp.
  371--373, 2004.

\bibitem{Kish2018}
F.~Kish, V.~Lal, P.~Evans, S.~W. Corzine, M.~Ziari, T.~Butrie, M.~Reffle, H.~S.
  Tsai, A.~Dentai, J.~Pleumeekers, M.~Missey, M.~Fisher, S.~Murthy,
  R.~Salvatore, P.~Samra, S.~Demars, N.~Kim, A.~James, A.~Hosseini,
  P.~Studenkov, M.~Lauermann, R.~Going, M.~Lu, J.~Zhang, J.~Tang, J.~Bostak,
  T.~Vallaitis, M.~Kuntz, D.~Pavinski, A.~Karanicolas, B.~Behnia, D.~Engel,
  O.~Khayam, N.~Modi, M.~R. Chitgarha, P.~Mertz, W.~Ko, R.~Maher, J.~Osenbach,
  J.~T. Rahn, H.~Sun, K.~T. Wu, M.~Mitchell, and D.~Welch, ``{System-on-Chip
  Photonic Integrated Circuits},'' \emph{IEEE Journal of Selected Topics in
  Quantum Electronics}, vol.~24, no.~1, 2018.

\bibitem{Smit2014}
M.~Smit, X.~Leijtens, H.~Ambrosius, E.~Bente, J.~{Van Der Tol}, B.~Smalbrugge,
  T.~{De Vries}, E.~J. Geluk, J.~Bolk, R.~{Van Veldhoven}, L.~Augustin,
  P.~Thijs, D.~D'Agostino, H.~Rabbani, K.~Lawniczuk, S.~Stopinski, S.~Tahvili,
  A.~Corradi, E.~Kleijn, D.~Dzibrou, M.~Felicetti, E.~Bitincka, V.~Moskalenko,
  J.~Zhao, R.~Santos, G.~Gilardi, W.~Yao, K.~Williams, P.~Stabile,
  P.~Kuindersma, J.~Pello, S.~Bhat, Y.~Jiao, D.~Heiss, G.~Roelkens, M.~Wale,
  P.~Firth, F.~Soares, N.~Grote, M.~Schell, H.~Debregeas, M.~Achouche, J.~L.
  Gentner, A.~Bakker, T.~Korthorst, D.~Gallagher, A.~Dabbs, A.~Melloni,
  F.~Morichetti, D.~Melati, A.~Wonfor, R.~Penty, R.~Broeke, B.~Musk, and
  D.~Robbins, ``{An introduction to InP-based generic integration
  technology},'' \emph{Semiconductor Science and Technology}, vol.~29, no.~8,
  2014.

\bibitem{Smit2019}
M.~Smit, K.~Williams, and J.~{Van Der Tol}, ``{Past, present, and future of
  InP-based photonic integration},'' \emph{APL Photonics}, vol.~4, no.~5, 2019.

\bibitem{Osowski1997}
M.~L. Osowski, R.~M. Lammert, and J.~J. Coleman, ``{A dual-wavelength source
  with monolithically integrated electroabsorption modulators and Y-junction
  coupler by selective-area MOCVD},'' \emph{IEEE Photonics Technology Letters},
  vol.~9, no.~2, pp. 158--160, 1997.

\bibitem{Teng2000}
J.~H. Teng, S.~J. Chua, Z.~H. Zhang, Y.~H. Huang, G.~Li, and Z.~J. Wang,
  ``{Dual-Wavelength Laser Source Monolithically Integrated with Y-Junction
  Coupler and Isolator Using Quantum-Well Intermixing},'' \emph{IEEE Photonics
  Technology Letters}, vol.~12, no.~10, pp. 1310--1312, 2000.

\bibitem{Sun2016}
M.~Sun, S.~Tan, S.~Liu, F.~Guo, D.~Lu, R.~Zhang, Q.~Kan, and C.~Ji,
  ``{Monolithically integrated two-wavelength distributed Bragg reflector laser
  for terahertz generation},'' \emph{2016 Conference on Lasers and
  Electro-Optics, CLEO 2016}, pp. 4--5, 2016.

\bibitem{Zhao2018}
D.~Zhao, S.~Andreou, W.~Yao, D.~Lenstra, K.~Williams, and X.~Leijtens,
  ``{Monolithically integrated multiwavelength laser with optical
  feedback:Damped relaxation oscillation dynamics and narrowed linewidth},''
  \emph{IEEE Photonics Journal}, vol.~10, no.~6, pp. 1--8, 2018.

\bibitem{Guzman2021}
R.~Guzman, G.~Carpintero, M.~Ali, J.~Cesar, A.~Zarzuelo, L.~Gonzalez-Guerrero,
  C.~G. Roeloffzen, R.~Grootjans, J.~P. Epping, and I.~Visscher, ``{Widely
  Tunable RF Signal Generation Using an InP/Si3N4 Hybrid Integrated
  Dual-Wavelength Optical Heterodyne Source},'' \emph{Journal of Lightwave
  Technology}, vol.~39, no.~24, pp. 7664--7671, 2021.

\bibitem{Gioannini2014a}
M.~Gioannini, M.~Dommermuth, L.~Drzewietzki, I.~Krestnikov, D.~Livshits,
  M.~Krakowski, and S.~Breuer, ``{Two-state semiconductor laser self-mixing
  velocimetry exploiting coupled quantum-dot emission-states: experiment,
  simulation and theory},'' \emph{Optics Express}, vol.~22, no.~19, p. 23402,
  2014.

\bibitem{Qi2011}
\BIBentryALTinterwordspacing
X.~Q. Qi and J.~M. Liu, ``{Photonic microwave applications of the dynamics of
  semiconductor lasers},'' \emph{IEEE Journal on Selected Topics in Quantum
  Electronics}, vol.~17, no.~5, pp. 1198--1211, 2011. [Online]. Available:
  \url{http://ieeexplore.ieee.org.ezp.lib.rochester.edu/abstract/document/5749684/}
\BIBentrySTDinterwordspacing

\bibitem{Docter2010}
B.~Docter, J.~Pozo, S.~Beri, I.~V. Ermakov, J.~Danckaert, M.~K. Smit, and
  F.~Karouta, ``{Discretely tunable laser based on filtered feedback for
  telecommunication applications},'' \emph{IEEE Journal on Selected Topics in
  Quantum Electronics}, vol.~16, no.~5, pp. 1405--1412, 2010.

\bibitem{Ermakov2012}
I.~V. Ermakov, S.~Beri, M.~Ashour, J.~Danckaert, B.~Docter, J.~Bolk, X.~J.
  Leijtens, and G.~Verschaffelt, ``{Semiconductor ring laser with on-chip
  filtered optical feedback for discrete wavelength tuning},'' \emph{IEEE
  Journal of Quantum Electronics}, vol.~48, no.~2, pp. 129--136, 2012.

\bibitem{Khoder2013a}
M.~Khoder, G.~Verschaffelt, R.~M. Nguimdo, X.~J.~M. Leijtens, J.~Bolk, and
  J.~Danckaert, ``{Switchable multiwavelength emission using semiconductor ring
  laser with optical filtered feedback},'' \emph{2013 Conference on Lasers {\&}
  Electro-Optics Europe {\&} International Quantum Electronics Conference CLEO
  EUROPE/IQEC}, vol.~38, no.~14, pp. 1--1, 2013.

\bibitem{Khoder2013b}
M.~Khoder, G.~Verschaffelt, R.~M. Nguimdo, X.~Leijtens, J.~Bolk, and
  J.~Danckaert, ``{Controlled multiwavelength emission using semiconductor ring
  lasers with on-chip filtered optical feedback},'' \emph{Opt. Lett.}, vol.~38,
  no.~14, pp. 2608--2610, 2013.

\bibitem{Carpintero2012}
G.~Carpintero, E.~Rouvalis, K.~{\L}awniczuk, M.~Fice, C.~C. Renaud, X.~J.~M.
  Leijtens, E.~A. J.~M. Bente, M.~Chitoui, F.~{Van Dijk}, and A.~J. Seeds,
  ``{95 GHz millimeter wave signal generation using an arrayed waveguide
  grating dual wavelength semiconductor laser},'' \emph{Optics Letters},
  vol.~37, no.~17, p. 3657, 2012.

\bibitem{Markus2003a}
A.~Markus, J.~X. Chen, C.~Parantho{\"{e}}n, A.~Fiore, C.~Platz, and
  O.~Gauthier-Lafaye, ``{Simultaneous two-state lasing in quantum-dot
  lasers},'' \emph{Applied Physics Letters}, vol.~82, no.~12, pp. 1818--1820,
  2003.

\bibitem{Naderi2010a}
N.~A. Naderi, F.~Grillot, K.~Yang, J.~B. Wright, A.~Gin, and L.~F. Lester,
  ``{Two-color multi-section quantum dot distributed feedback laser},''
  \emph{Optics Express}, vol.~18, no.~26, p. 27028, 2010.

\bibitem{Wang2007}
H.-Y. Wang, H.-C. Cheng, S.-D. Lin, and C.-P. Lee, ``{Wavelength switching
  transition in quantum dot lasers},'' \emph{Applied Physics Letters}, vol.~90,
  no.~8, p. 081112, 2007.

\bibitem{Breuer2013}
S.~Breuer, M.~Rossetti, L.~Drzewietzki, I.~Montrosset, M.~Krakowski,
  M.~Hopkinson, and W.~Els{\"{a}}sser, ``{Dual-state absorber-photocurrent
  characteristics and bistability of two-section quantum-dot lasers},''
  \emph{IEEE Journal on Selected Topics in Quantum Electronics}, vol.~19,
  no.~5, pp. 1--9, 2013.

\bibitem{Roh2000}
S.~D. Roh, S.~Member, T.~S. Yeoh, S.~Member, R.~B. Swint, S.~Member, A.~E.
  Huber, C.~Y. Woo, J.~S. Hughes, and J.~J. Coleman, ``{Dual-Wavelength
  InGaAs-GaAs Ridge Waveguide Distributed Bragg Reflector Lasers with Tunable
  Mode Separation},'' \emph{IEEE Photonics Technology Letters}, vol.~12,
  no.~10, pp. 1307--1309, 2000.

\bibitem{Gwaro2020}
J.~O. Gwaro, C.~Brenner, L.~S. Theurer, M.~Maiwald, B.~Sumpf, and M.~R.
  Hofmann, ``{Continuous Wave THz System Based on an Electrically Tunable
  Monolithic Dual Wavelength Y-Branch DBR Diode Laser},'' \emph{Journal of
  Infrared, Millimeter, and Terahertz Waves}, vol.~41, no.~5, pp. 568--575,
  2020.

\bibitem{Mak2021}
J.~Mak, A.~van Rees, R.~E.~M. Lammerink, D.~Geskus, Y.~Fan, P.~J.~M. van~der
  Slot, C.~G.~H. Roeloffzen, and K.-J. Boller, ``{High spectral purity
  microwave generation using a dual-frequency hybrid integrated
  semiconductor-dielectric waveguide laser},'' \emph{OSA Continuum}, vol.~4,
  no.~8, p. 2133, 2021.

\bibitem{Huang2021}
X.~Huang, C.~R. Doerr, C.~Qin, J.~Heanue, N.~Zhu, D.~Ton, B.~Guan, S.~Zhang,
  and Y.~Zhao, ``{Silicon-photonic laser emitting tunable dual wavelengths with
  highly correlated phase noise},'' \emph{Optics Letters}, vol.~46, no.~1, p.
  142, 2021.

\bibitem{Koester2022}
J.~P. Koester, H.~Wenzel, J.~Fricke, O.~Brox, A.~Zeghuzi, A.~Muller, L.~S.
  Theurer, B.~Sumpf, A.~Knigge, and G.~Trankle, ``{Comparative Study of
  Monolithic Integrated MMI-Coupler-Based Dual-Wavelength Lasers},'' \emph{IEEE
  Journal of Selected Topics in Quantum Electronics}, vol.~28, no.~1, 2022.

\bibitem{Osborne2007}
S.~Osborne, S.~O'Brien, K.~Buckley, R.~Fehse, A.~Amann, J.~Patchell, B.~Kelly,
  D.~R. Jones, J.~O'Gorman, and E.~P. O'Reilly, ``{Design of single-mode and
  two-color fabry - p{\'{e}}rot lasers with patterned refractive index},''
  \emph{IEEE Journal on Selected Topics in Quantum Electronics}, vol.~13,
  no.~5, pp. 1157--1163, 2007.

\bibitem{Zhang2022}
\BIBentryALTinterwordspacing
Y.~Zhang, B.~Yuan, J.~Shi, W.~Qi, L.~Li, L.~Wang, J.~Zheng, S.~Guan, T.~Fang,
  and X.~Chen, ``{A Stable Dual-Wavelength DFB Semiconductor Laser With
  Equivalent Chirped Sampled Grating},'' \emph{IEEE Journal of Quantum
  Electronics}, vol.~58, no.~1, pp. 1--7, feb 2022. [Online]. Available:
  \url{https://ieeexplore.ieee.org/document/9606696/}
\BIBentrySTDinterwordspacing

\bibitem{Osborne2012a}
S.~Osborne, P.~Heinricht, N.~Brandonisio, A.~Amann, and S.~Obrien,
  ``{Wavelength switching dynamics of two-colour semiconductor lasers with
  optical injection and feedback},'' \emph{Semiconductor Science and
  Technology}, vol.~27, no.~9, 2012.

\bibitem{Virte2014}
M.~Virte, S.~Breuer, M.~Sciamanna, and K.~Panajotov, ``{Switching between
  ground and excited states by optical feedback in a quantum dot laser
  diode},'' \emph{Applied Physics Letters}, vol. 105, no.~12, p. 121109, 2014.

\bibitem{Virte2016}
M.~Virte, R.~Pawlus, M.~Sciamanna, K.~Panajotov, and S.~Breuer, ``{Energy
  exchange between modes in a multimode two-color quantum dot laser with
  optical feedback},'' \emph{Optics Letters}, vol.~41, no.~14, p. 3205, 2016.

\bibitem{Ohtsubo2013}
{Junji Ohtsubo}, \emph{{Semiconductor Lasers: Stability, Instability and
  Chaos}}, 3rd~ed.\hskip 1em plus 0.5em minus 0.4em\relax Springer-Verlag
  Berlin Heidelberg, 2013.

\bibitem{Koryukin2004a}
I.~V. Koryukin and P.~Mandel, ``{Dynamics of semiconductor lasers with optical
  feedback: Comparison of multimode models in the low-frequency fluctuation
  regime},'' \emph{Physical Review A - Atomic, Molecular, and Optical Physics},
  vol.~70, no. 5 B, pp. 1--6, 2004.

\end{thebibliography}

\end{document}